\documentclass[conference]{IEEEtran}
\IEEEoverridecommandlockouts
\usepackage{cite}
\usepackage{caption}
\usepackage{amsmath,amssymb,amsfonts}
\usepackage{algorithmic}
\usepackage{graphicx}
\usepackage{textcomp}
\usepackage{xcolor}
\usepackage{hyperref}
\def\BibTeX{{\rm B\kern-.05em{\sc i\kern-.025em b}\kern-.08em
    T\kern-.1667em\lower.7ex\hbox{E}\kern-.125emX}}
\begin{document}

\title{Proactive Market Making and Liquidity Analysis for Everlasting Options in DeFi Ecosystems}

\author{\IEEEauthorblockN{Hardhik Mohanty}
\IEEEauthorblockA{\textit{Viterbi School of Engineering} \\
\textit{University of Southern California}\\
Los Angeles, California \\
hmohanty@usc.edu}
\and
\IEEEauthorblockN{Giovanni Zaarour}
\IEEEauthorblockA{\textit{Viterbi School of Engineering} \\
\textit{University of Southern California}\\
Los Angeles, California \\
gzaarour@usc.edu}
\and
\IEEEauthorblockN{Bhaskar Krishnamachari}
\IEEEauthorblockA{\textit{Viterbi School of Engineering} \\
\textit{University of Southern California}\\
Los Angeles, California \\
bkrishna@usc.edu}
}

\maketitle

\begin{abstract}
Everlasting options, a relatively new class of perpetual financial derivatives, have emerged to tackle the challenges of rolling contracts and liquidity fragmentation in decentralized finance markets. This paper offers an in-depth analysis of markets for everlasting options, modeled using a dynamic proactive market maker. We examine the behavior of funding fees and transaction costs across varying liquidity conditions. Using simulations and modeling, we demonstrate that liquidity providers can aim to achieve a net positive PnL by employing effective hedging strategies, even in challenging environments characterized by low liquidity and high transaction costs. Additionally, we provide insights into the incentives that drive liquidity providers to support the growth of everlasting option markets and highlight the significant benefits these instruments offer to traders as a reliable and efficient financial tool.
\end{abstract}


\section{Introduction}

Everlasting options are a novel class of perpetual financial derivatives that allow holders to maintain their positions indefinitely by paying a daily funding fee. This funding fee mechanism eliminates the need to roll over contracts, which is a common practice in traditional fixed-term options \cite{singh2024option}. Rolling over fixed-term options involves selling an expiring option and purchasing a new one with a long expiration. Such a process incurs significant transaction costs, including a spread paid to the market makers. These cumulative costs and the liquidity fragmentation caused by maintaining separate markets for different expiration dates make everlasting options a more efficient and attractive alternative.

One of the key advantages of everlasting options is that their liquidity is concentrated in a single, unified market, which reduces liquidity fragmentation. The funding fee for these options is determined daily as the difference between the market price of the option and its payoff at the end of the day. As demonstrated in \cite{white2019everlasting}, the valuation of an everlasting option can be equivalently represented as a decreasing weighted sum of a portfolio of fixed-term options with consecutive expiration dates, leveraging established pricing models such as the Black-Scholes framework \cite{black1973pricing}. 

\begin{table}[!t]
\centering
\scriptsize 
\setlength{\tabcolsep}{2pt} 
\renewcommand{\arraystretch}{1.2} 
\caption{DeFi Platforms Offering Options}
\begin{tabular}{|p{1.8cm}|p{1.8cm}|p{1.8cm}|p{2.5cm}|}
\hline
\textbf{Platform} & \textbf{Total Liquidity} & \textbf{Trading Volume} & \textbf{Market Maker} \\ \hline
Hegic & $\sim\$7$ Million & $\sim\$1$ Billion & Internal AMM \\ \hline
Opyn & $\sim\$40$ Million & $\sim\$500$ Million & External Liquidity Providers \\ \hline
Deri Protocol & $\sim\$30$ Million & $\sim\$70$ Million & Proactive Market Maker (PMM) \\ \hline
Derive & $\sim\$33$ Million & $\sim\$4$ Billion & Internal AMM \& Synth Market Makers \\ \hline
dYdX & $\sim\$428$ Million & $\sim\$274$ Billion & Order Book Model \\ \hline
GammaSwap & $\sim\$14$ Million & $\sim\$400$ Million & Internal \& External AMM \\ \hline
\end{tabular}
\vspace{0.5em}
\label{table:defi_options}
\end{table}


This paper builds on the foundational concepts of everlasting options and investigates their potential to transform derivative markets in DeFi. In particular, our contributions are listed as follows:
\begin{itemize}
    \item We provide a detailed analysis of funding fees and transaction costs under varying liquidity conditions, offering insights into market dynamics.
    \item We propose and evaluate a $\Delta$-hedging strategy for Liquidity Providers (LPs), demonstrating the possibility of profitability even in low-liquidity and high transaction-cost markets.
    \item We explore the dynamics of different market making strategies for options, highlighting the role of incentives and LP participation in enhancing market efficiency.
\end{itemize}


\section{Related Work}\label{sec:related_work}


The introduction of Bitcoin perpetual swaps by BitMEX \cite{alexander2020bitmex} in 2016 was a pivotal moment, introducing cryptocurrency derivatives. Derivatives allow traders to gain exposure to cryptocurrencies without holding the underlying asset, facilitating strategic portfolio management and market liquidity. Options are among the most widely used derivatives, providing traders with the right, but not the obligation, to buy or sell an asset at a predetermined price before a specific expiration date \cite{hull1993options}. Despite their utility, traditional options markets in DeFi face challenges such as liquidity fragmentation and the costs associated with rolling over contracts as they near expiration. 


Perpetual futures are closely related to traditional futures contracts but differ in two significant ways: 1) They have no expiration date, and 2) They rely on a funding mechanism to maintain price parity with the underlying asset \cite{atzberger2024low}. 
Everlasting options, introduced in \cite{white2019everlasting}, reduce the need to roll over positions, thus minimizing transaction costs and liquidity fragmentation. Everlasting options are priced according to their equivalence to a portfolio of fixed-term options with diminishing weights, taking advantage of established valuation models such as Black-Scholes \cite{black1973pricing}. This innovative approach provides traders with continuous exposure while simplifying market operations and improving liquidity efficiency \cite{madrigal2022time}. Unlike traditional automated market makers (AMMs) that rely on external arbitrage mechanisms, Dynamic Proactive Market Makers (DPMMs), discussed in~\cite{alpha2021exchange}, are a more efficient mechanism for price discovery, rapidly aligning market prices with external reference prices, making them particularly suitable for everlasting options.

Existing studies have explored various aspects of liquidity provisioning, market-making strategies, and financial risks in decentralized finance \cite{bichuch2024defi}, \cite{xu2025improving}, \cite{qin2021empirical}. 
While these works contribute significantly to understanding liquidity provisioning, risk management, and financial stability in DeFi, they primarily focus on traditional AMMs, lending protocols, and concentrated liquidity mechanisms. To the best of our knowledge, our work is the first to conduct a comprehensive liquidity analysis specifically for everlasting options, a novel derivative instrument in DeFi. 
Liquidity is a key factor in the scaling of DeFi derivatives platforms. Table-\ref{table:defi_options} presents the overview of the most popular DeFi platforms offering options contracts and futures.

\begin{table}[!t]
\scriptsize 
\setlength{\tabcolsep}{4pt} 
\renewcommand{\arraystretch}{1.2} 
\centering
\caption{Notations and Descriptions}
\begin{tabular}{|p{1.6cm}|p{5.8cm}|} 
\hline
\textbf{Notation} & \textbf{Description} \\ \hline
$S_t$ & Price of the underlying asset at time $t$ \\ \hline
$\mu$ & Annualized drift of the underlying asset \\ \hline
$\sigma$ & Annualized volatility of the underlying asset \\ \hline
$r$ & Risk-free rate \\ \hline
$T$ & Time to maturity \\ \hline
$\Phi(\cdot)$ & CDF of the standard normal distribution \\ \hline
$P_m$ & Adjusted mark price of the option \\ \hline
$Q_0$ & Reference liquidity in the liquidity pool \\ \hline
$V$ & Inventory of options held by the liquidity pool \\ \hline
$k$ & Shape parameter for the dynamic proactive market maker \\ \hline
$F_t$ & Funding fee on day $t$ \\ \hline
$G_c$ & Transaction cost \\ \hline
$G_b$ & Fixed gas fee per transaction \\ \hline
$\Delta$ & Sensitivity to the underlying asset's price \\ \hline
$h$ & Hedge ratio \\ \hline
$\Pi_t$ & Position in the underlying asset for hedging \\ \hline
\end{tabular}
\vspace{0.5em}
\label{table:notations}
\end{table}

\section{Methodology}\label{sec:methodology}

This section outlines the methodological approach taken to simulate the liquidity pool for everlasting options \footnote{Github link: \href{https://github.com/ANRGUSC/liquidity-everlasting-options}{https://github.com/ANRGUSC/liquidity-everlasting-options}}. 
It is organized into six main parts. Table-\ref{table:notations} describes all the notations used in this section.

\begin{figure}[!t]
    \centering
    \includegraphics[width=0.9\columnwidth]{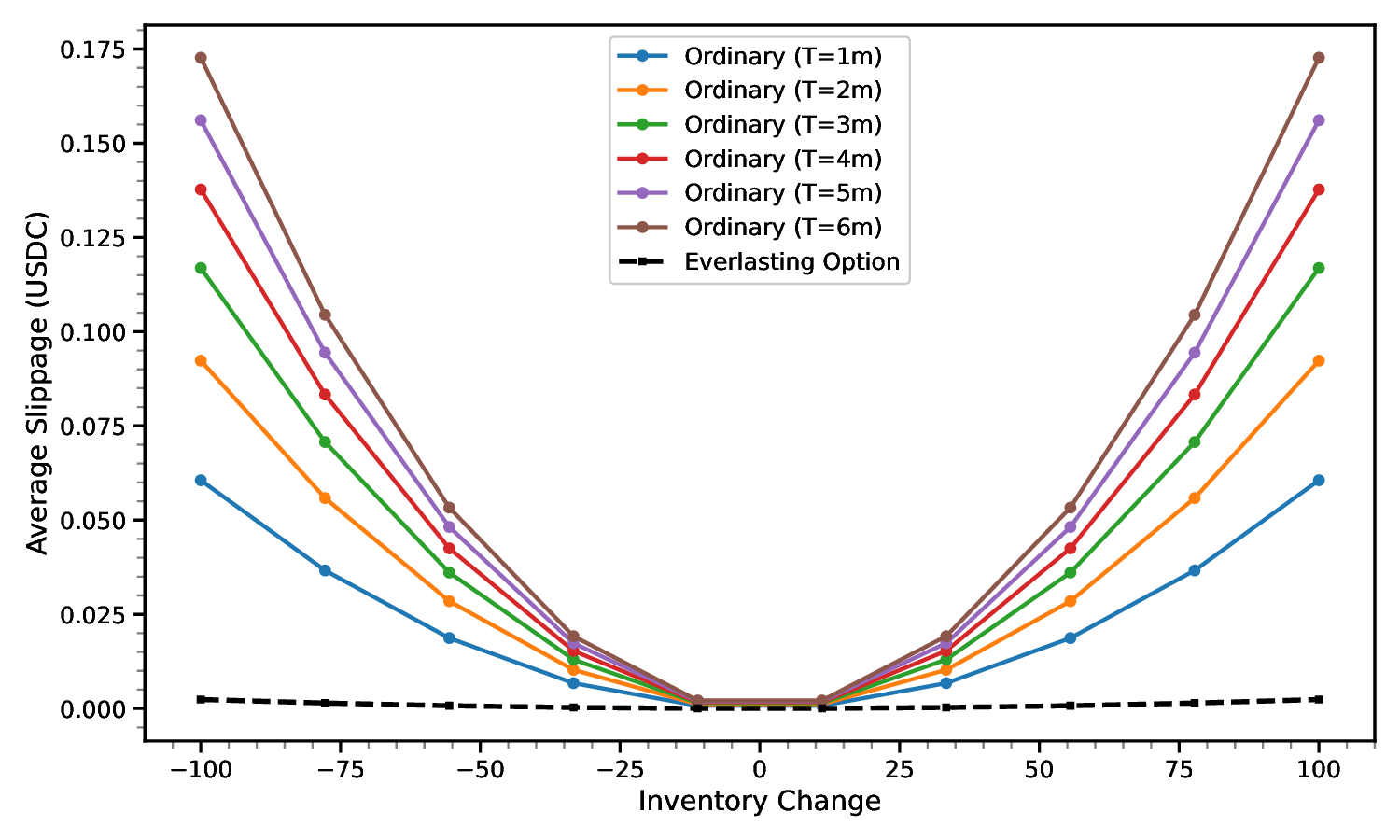} 
    \caption{Slippage: Everlasting vs. Fixed-Expiration Options}
    \label{fig:slippage_demo}
\end{figure}

\begin{figure*}[!t]
    \centering
    \includegraphics[width=\textwidth]{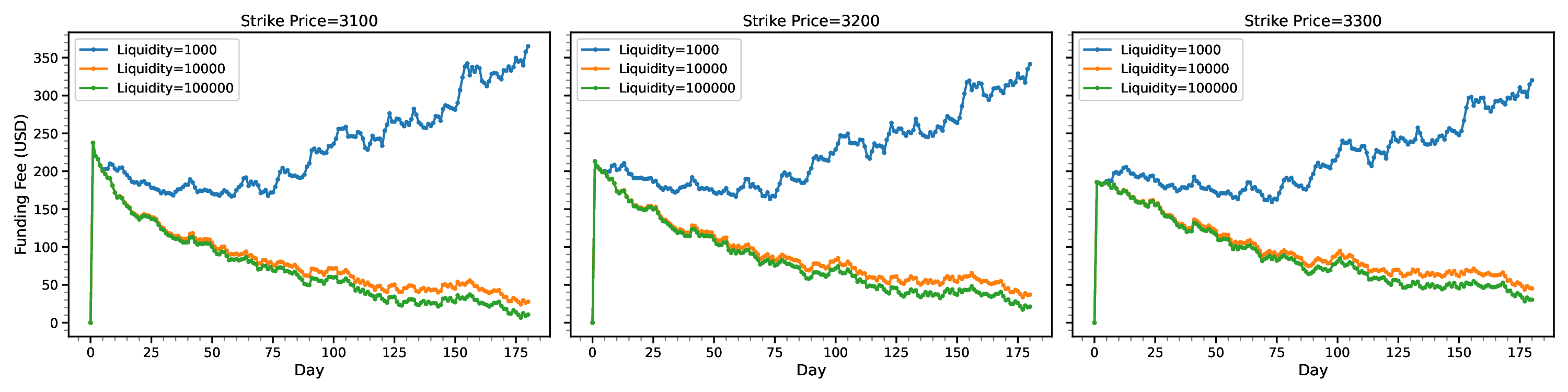}
    \caption{Behavior of funding fees with different liquidity levels across different strikes}
    \label{fig:funding_fees}
\end{figure*}

\subsection{Geometric Brownian Motion (GBM) Path Generation}
The underlying asset price $S_t$ is assumed to follow a Geometric Brownian Motion (GBM) \cite{duffie2010dynamic}, which we discretize into daily steps
$\Delta t = 1/365$. It has two key properties: (1) lognormal distribution of prices, which ensures that the asset price remains strictly positive, and (2) continuously compounded returns that follow a normal distribution. Over each day,
\begin{equation}
S_{t+\Delta t}=S_t \exp \left(\left(\mu-\frac{1}{2} \sigma^2\right) \Delta t+\sigma \sqrt{\Delta t} Z\right)
\end{equation}
where, $Z \sim \mathcal{N}(0,1)$. This process is repeated over a predefined horizon (180 days), and multiple independent paths can be generated by sampling new standard normals $Z$. Each path starts at an initial price $S_0$. 


\subsection{Price Calculation of Everlasting Option}
We can use the standard Black-Scholes formula \cite{black1973pricing} to price fixed-expiration European call options. For a call option with strike $K$, the pricing formula is as follows:
\begin{equation}
C(S, K, r, \sigma, T)=S \Phi\left(d_1\right)-K e^{-r T} \Phi\left(d_2\right)
\end{equation}
where,
\begin{equation}
d_1=\frac{\ln (S / K)+\left(r+\frac{1}{2} \sigma^2\right) T}{\sigma \sqrt{T}}, \quad d_2=d_1-\sigma \sqrt{T}
\end{equation}
Now to approximate the price of an everlasting call option, we can employ a weighted sum of fixed-expiration calls. For maturities $T_i$, each call option's price is discounted accordingly and then summed:
\begin{equation}
C_\text { EO } \approx \frac{1}{D} \sum_{i=1}^n\left(\frac{D}{D+1}\right)^i\operatorname{C}\left(S, K, r, \sigma, T_i\right)\
\end{equation}
where $D$ is a chosen integer factor controlling the decay and $n$ is intended to capture the time horizon.

\subsection{Dynamic Proactive Market Making}

We adopt the ideas from \cite{alpha2021exchange} to simulate a dynamic proactive market maker (DPMM). Let $i_{value}$ denote the theoretical or fair value of the derivative. A k-curve DPMM updates the quoted price depending on the current inventory as:
\begin{equation}
\text{P}_{m} = i_{\text{value}}\left(1+k\left(\frac{V}{Q_0}\right)^2\right)
\end{equation}
A higher inventory leads to a proportionally higher (or lower) mark price ($P_m$). In the case of fixed-expiration options, the liquidity is fragmented across different pools corresponding to the expiration dates. All these liquidity pools apply a similar adjustment mechanism but scaled to its respective liquidity and inventory size.

\subsection{Funding Fees and Transaction Costs}
At each daily step, the difference between the updated mark price and the immediate payoff of the option is calculated as the funding fee. If $P^t_m$ is the new mark price on day $t$ and $\text{payoff}_t = max(S_t - K, 0)$,
\begin{equation}
    F_t = P^t_m - \text{payoff}_t
\end{equation}
where, positive funding fees ($F_t > 0$) indicates that longs pay to shorts and negative funding fees ($F_t < 0$) indicates shorts pay longs.
Also, each time the LP's inventory changes by some amount, there is a cost associated with the transaction, which includes a base gas fee and additional dynamic components that account for the pool's liquidity, transaction volume, market volatility, and network congestion \cite{kim2024optimal}. If the pool quotes a new mark price, the transaction cost is calculated as follows:

\begin{align}
\mathcal{G}_c &= \mathcal{G}_b
+ \left(i_m \cdot 
\frac{Q_0}{Q_0 + V \cdot (1 - \sigma) \cdot N)}\right) \cdot \left(1 + \eta \cdot U\right)
\end{align}
where, $i_m$ reflects the proportional impact cost factor, $\eta$ denotes the network congestion level, $N \sim \mathcal{N}(0, 1)$ refers to the standard normal distribution, and $U \sim \mathcal{U}(0, 1)$ denotes the uniform random variable. This formula provides a comprehensive model for real-world costs of on-chain transactions, incorporating dynamic factors to reflect the liquidity pool's conditions and market environment.

\begin{figure*}[!t]
    \centering
    \includegraphics[width=\textwidth]{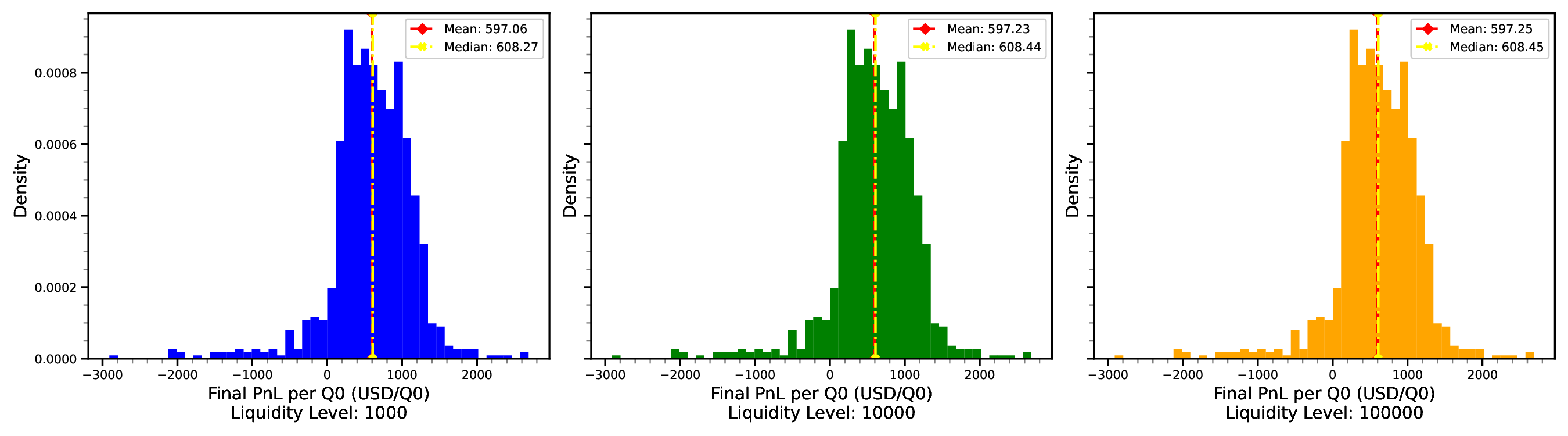}
    \caption{Histograms of Final PnL (Normalized by $Q_0$) for Different Liquidity Levels.}
    \label{fig:histograms}
\end{figure*}

\subsection{$\Delta$-Hedging by Liquidity Providers}
The LP plays a critical role in the everlasting options market by providing liquidity (typically in the form of USDC) to act as the counterparty for option contracts. This means that the LP enables traders to buy or sell options by maintaining a reserve of capital (referred to as the inventory). The inventory represents the pool of options held by the LP, which can fluctuate depending on market activity.
As a result of providing liquidity, the LP accumulates a net short or long exposure to the options, which can be highly sensitive to the underlying asset's price movements. To manage this risk, the LP implements $\Delta$-hedging \cite{matic2023hedging}. This strategy involves offsetting the LP's exposure to price movements of the underlying asset by taking a counter-position in the asset itself \cite{lyra2021whitepaper}. For a call option, the delta is approximated as $\Delta \approx \Phi(d_1)$. Therefore, the LP's net delta on day $t$ is given as:
\begin{equation}
    \Delta^{net}_t = V_t \cdot \Delta(S_t,K,r,\sigma,T)
\end{equation}
For an everlasting option, we can approximate $\Delta$ by using several fixed-expiration maturities. The hedge ratio $h$ controls the fraction of this exposure the LP wishes to offset. The LP holds a position ($\Pi$) in the underlying asset such that:
\begin{equation}
    \Pi_t = -h \cdot \Delta^{net}_t
\end{equation}
Rebalancing occurs daily as the inventory or underlying asset price changes. An inventory cap is also imposed so that the pool does not accept unbounded exposure, reducing tail risks.

\subsection{PnL Calculation}
On each day $t$, the LP's profit and loss (PnL) has three components:
\begin{equation}
    \text{PnL}_t = \text{PnL}^\text{hedge}_t + F_t - \mathcal{G}_c
\end{equation}
The LP's cumulative PnL is calculated by summing the daily PnL over the trading horizon. 

\section{Simulation Parameters}\label{sec:simulation_parameters}

We simulate the Ethereum (ETH) market \cite{buterin2014next} to evaluate liquidity pools for everlasting options. The simulations are designed to reflect realistic market conditions and operational dynamics of ETH-based derivatives.
The initial price of ETH is set \$3000, which corresponds to ETH's historical price range during periods of moderate market activity. 
The annual drift rate is set to 3\% to represent an estimate of ETH's expected return over the simulation period. An annual volatility of 60\% reflects ETH's historical price fluctuations, capturing the asset's inherent high-risk profile. The simulation horizon spans 6 months, sufficient to observe significant price movements and impact on LP's performance. We conduct 100 independent simulation runs to capture a diverse array of market scenarios.


\section{Experimental Results}\label{sec:experimental_results}
\subsection{Slippage Analysis}
Figure-\ref{fig:slippage_demo} demonstrates the slippage characteristics of everlasting options and fixed-expiration options with varying inventory changes. The analysis was conducted using a DPMM model, where liquidity was divided between a single unified everlasting option pool and 6 separate fixed-expiration option pools corresponding to expiration dates ranging from 1 to 6 months. The results show that slippage for everlasting options remains nearly constant across the entire range of inventory changes due to high unified liquidity concentration. 
It shows that everlasting options can help reduce trading costs, particularly for high-volume trading scenarios where inventory changes are significant. In contrast, fixed-expiration options exhibit a parabolic increase in slippage as the inventory increases or decreases. 

\subsection{Funding Fees \& Transaction Costs}
The funding fee behavior was analyzed for different liquidity levels $Q_0 \in \{1000, 10000, 100000\}$ USDC and strike prices $K \in \{3100, 3200, 3300\}$ using a momentum-based simulation approach. Figure-\ref{fig:funding_fees} depicts the temporal evolution of funding fees over 180 days. The experiment results indicate that higher liquidity results in lower funding fees and less volatility, and higher strike prices result in larger funding fees.


\subsection{LP Profitability Analysis}

The histograms in Figure-\ref{fig:histograms} illustrate the distribution of final Profit and Loss (PnL) normalized by liquidity for three different scales $Q_0 \in \{1000, 10000, 100000\}$ USDC (left to right). Across all liquidity levels, the distributions show a Gaussian-like shape centered around positive PnL values. This indicates that the DPMM combined with $\Delta$-hedging strategy successfully minimizes tail risks while maintaining profitability. 
As $Q_0$ increases, both the mean and median PnL values improve, as depicted by the vertical dashed lines in each histogram plot. 
To further support our claims, we present results from real-world financial data for ETH prices and everlasting options, demonstrating how our $\Delta$-hedging strategy performs under practical scenarios. The liquidity provider hedges their exposure with an everlasting call option with strike = \$3200 by dynamically adjusting their ETH holdings based on $\Delta$. Our results show a final PnL of +\$220.74.

\subsection{Comparison of AMM vs DPMM}

We analyze LP profitability in different market making scenarios by simulating both AMM and DPMM models. Our findings indicate that the DPMM model achieved a mean PnL of approximately \$349K, a median of \$153K, and a volatility of about \$555K, which yields a Sharpe ratio of 0.63, a win rate of 90.8\%, and a profit factor of 22.4. In contrast, the AMM model produced a higher mean PnL near \$984K but a slightly negative median of roughly -\$11K, with a much higher volatility of approximately \$2.28M, a Sharpe ratio of 0.43, a win rate of 48.6\%, and a profit factor of 11.4. The robust performance of the DPMM can be primarily attributed to its inventory-sensitive pricing and active hedging strategies. In contrast, the static pricing mechanism of AMM can lead to larger deviations during market swings.

\section{Conclusions and Future Works}\label{sec:conclusion}
We presented a comprehensive analysis of liquidity pools for everlasting options. Using the DPMM framework we have demonstrated that LPs can aim to achieve net positive profitability through an effective hedging strategy even in scenarios characterized by low liquidity and high transaction costs. We observed that everlasting options offer a significant improvement over fixed-expiration options by consolidating liquidity into a single pool. The funding fee mechanism, when paired with adequate liquidity levels, ensures stability and reduces volatility for both traders and LPs. In addition, we saw that LPs employing $\Delta$ -hedging can effectively mitigate risks, with profitability positively correlated with increased liquidity levels. These insights spotlight the potential of everlasting options to reshape derivative markets in DeFi, offering a robust and scalable alternative to traditional fixed-expiration derivatives.
Future work could explore investor perspectives on trading everlasting options, improve understanding of funding fees, and develop novel market-making strategies to improve risk management and profitability of options markets in DeFi.

\label{Bibliography}

\bibliographystyle{IEEEtran}
\bibliography{references}

\end{document}